\documentclass[12pt]{article}
\usepackage{amsmath,amsthm,amsfonts,graphicx,epsfig}
\usepackage[usenames]{color}
\numberwithin{equation}{section}
\newcommand{\mbf}[1]{\mathbf #1}

\newcommand{\e}{\varepsilon}
\newcommand{\dbar}{\kern-.1em{\raise.8ex\hbox{ -}}\kern-.6em{d}}

\def \be{\begin{equation}}
\def \ee{\end{equation}}
\def \bea{\begin{eqnarray}}
\def \eea{\end{eqnarray}}
\def \be{\begin{equation}}
\def \ee{\end{equation}}

\author{J. E.  Avron,  O. Raz
\\
Department of Physics\\ Technion, 32000 Haifa, Israel}

\begin{document}
\title{A geometric theory of swimming: Purcell's swimmer and its
symmetrized cousin}
\date{\today}%
\maketitle
\begin{abstract}
We develop a qualitative geometric approach to swimming at low Reynolds number which
avoids solving differential equations and uses instead landscape
figures of two notions of curvatures: The swimming curvature and
the curvature derived from dissipation. This approach gives
complete information for swimmers that swim on a line without
rotations and gives the main qualitative features for general
swimmers that can also rotate. We illustrate this
approach for a symmetric version of Purcell's swimmer which we
solve by elementary analytical means within slender body theory.
We then apply the theory to derive the basic qualitative
properties of Purcell's swimmer.
\end{abstract}
\section{Introduction}

\begin{figure}[htb]
  \includegraphics[width=8cm]{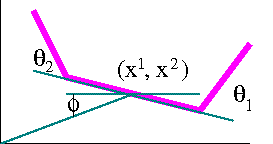}\\
  \caption{Purcell's swimmer.  The swimmer controls the angles $|\theta_{1,2}|
  <\pi$. 
  The location of the swimmer in the plane is determined by two position coordinates $x^{1,2}$ and
  one orientation  $x^0=\phi$.  The
  lengths of each arm is $\ell$ and the length of the body $\ell_0$.}\label{fig:purcell}
\end{figure}

Micro-swimmers are of general interest lately, motivated  by both
engineering and biological problems
\cite{Golestanian,N-Linked-Rods,Hosoi,JFM-Stone,Nature,Pnas-swimming-efficiency,
Spiroplasma,pushmepullyou,treadmilling,wada}. They can be
remarkably subtle as was illustrated by E. M. Purcell in his
famous talk on ``Life at low Reynolds numbers" \cite{Purcell}
where he introduced a deceptively simple swimmer shown in
Fig.~\ref{fig:purcell}. Purcell asked ``What will determine the
direction this swimmer will swim?'' This simple looking question
took 15 years to answer: Koehler, Becker and Stone
\cite{JFM-Stone} found that the direction of swimming depends,
among other things, on the stroke's {\em amplitude}: Increasing
the amplitudes of certain small strokes that propagate the swimmer
to the right result in propagation to the left. This shows that
even simple qualitative aspects of low Reynolds number swimming
can be quite un-intuitive.

Purcell's swimmer made of three slender rods  can be readily
analyzed numerically by solving three coupled, non-linear, first
order, differential equations \cite{Hosoi}.  However, at present
there appears to be no general method that can be used to gain
direct qualitative insight into the properties of the solutions of
these equations.

Our first aim here is to to describe a geometric approach which
allows one to describe the qualitative features of the solution of
the swimming differential equations without actually solving them.
Our tools are geometric.  The first tool is the notion of
curvature borrowed  from non-Abelian gauge theory
\cite{Wilczek-Shapere}. This curvature can be represented
graphically by landscape diagrams such as
Figs.~\ref{fig:OptStrkHH},\ref{fig:FphiEta=2},\ref{fig:PurcellCurvX}
which capture  the qualitative properties of general swimming
strokes. We have taken care not to assume any pre-existing
knowledge about gauge theory on part of the reader. Rather, we
have attempted to use swimming as a natural setting where one can
build and develop a picture of the notions of non-Abelian gauge
fields. Purcell's original question, ``What will determine the
direction this swimmer will swim?'' can often be answered by
simply looking at such landscape pictures.

Our second tool is a notion of metric and curvature associated
with the dissipation. The ``dissipation curvature'' can be
described as a landscape diagram and it gives information on the
geometry of ``shape space''.   This gives us useful geometric
tools that give qualitative information on the solutions of rather
complicated differential equations.

We  begin by illustrating these geometric methods for the
symmetric version of Purcell's swimmer, shown in
Fig.~\ref{fig:symmetric-p}. Symmetry  protects the swimmer against
rotations so it can only swim on a straight line. This makes it
simple to analyze by elementary analytical means. In particular,
it is possible to predict, using the landscape portraits of the
swimming curvature Fig.~\ref{fig:OptStrkHH}, which way it will
swim.  In this (Abelian) case the swimming curvature gives full
quantitative information on the swimmer. We then turn to the
non-Abelian case of the usual Purcell's swimmer which can also
rotate. There are now several notions of swimming curvatures: The
rotation curvature and the two translation curvature. The
translation curvatures are non-Abelian. This means that they give
precise information of small strokes but this information can not
be integrated to learn about large strokes. This can be viewed as
a failure of Stokes integration theorem.  Nevertheless, as we
shall explain, they do give lots of qualitative information about
large strokes as well.  

\section{The symmetric Purcell's swimmer}
\begin{figure}
  \includegraphics[width=8 cm]{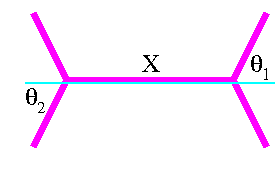}\\
  \caption{The symmetrized Purcell's swimmer can not rotate
  and can only move along the horizontal body-axis. The length of each arm is $\ell$
  and the length of the body $\ell_0$. The position of the swimmer is denoted by
  $X$. }\label{fig:symmetric-p}
\end{figure}

Purcell's swimmer, which was invented as ``The simplest animal
that can swim that way" \cite{Purcell}, is not simple to
analyze. A variant of it that is simple to analyze is shown in the
Fig ~\ref{fig:symmetric-p}.
The swimmer has four arms, each of length $\ell>0$ and one body
arm of length $\ell_0$, (possibly of zero length). The swimmer can
control the angles $\theta_j$ and the arms are not allowed to
touch. Both angles increase in the counterclockwise direction, $0<
\theta_{j}<\pi $.

Being symmetric, this swimmer can not rotate and can swim only in
the ``body'' direction. It falls into the class of ``simple
swimmers'' which includes the ``three linked spheres'' of Najafi
and Glolestanian \cite{Golestanian} and the Pushmepullyou
\cite{pushmepullyou}, whose hydrodynamics is elementary because
they can not turn.

\begin{figure}[htb]
  \includegraphics[width=12cm]{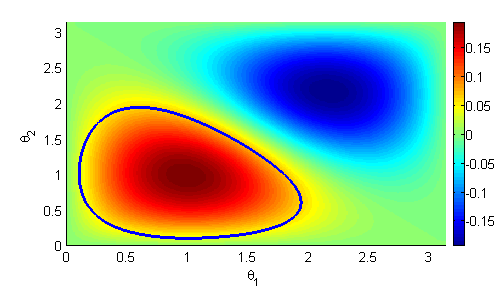}\\
  \caption{The swimming curvature (measured in the dissipation metric)
  on the surface of controls for the symmetric Purcell swimmer.
  The closed curve shows the optimal stroke. A bona fide optimizer
  exists since the
  swimming curvature vanishes on the boundary $\theta_j=\pi,0$   .}\label{fig:OptStrkHH}
\end{figure}

Let us first address Purcell's question ``What will determine the
direction this swimmer will swim?''  for the stroke shown in
Fig.~\ref{fig:OptStrkHH}.  In the stroke, the swimmer moves the
two arms backwards together and then bring them forward one by
one. The first half of the cycle pushes the swimmer forward and
the second half pulls it back. Which half of the cycle wins?

To answer that, one needs to remember that swimming at low
Reynolds numbers relies more on effective anchors than on good
propellers. Since one needs twice the force to drag a rod
transversally than to drag it along its axis
\cite{Happel-Brenner}, an open arm $\theta_j\approx \pi/2$ acts
like an anchor. This has the consequence that rowing with both
arms, in the same direction and in phase, is {\em less effective}
than bringing them back out of phase. The stroke actually swims
backwards.  This reasoning also shows that the swimmer is
Sisyphian: it performs a lot of forward and backward motion for
little net gain\footnote{Multimedia simulations can be viewed in
{\tt http://physics.technion.ac.il/\~{}avron}.}.

\subsection{The swimming equation}

The swimming equation at low Reynolds number is the requirement
that the total force (and torque) on the swimmer is zero. The
total force (and torque) is the sum of the forces (and torques) on
the four arms and body. For the symmetric swimmer, the torque and
force in the transversal direction vanish by symmetry. The
swimming equation is the condition that the force in the
body-direction vanishes. This force depends linearly on the known
rate of change of the controls, $\dot\theta$ and the unknown the
velocity $\dot X$ of the ``body-rod''. It gives a linear equation
for the velocity.

For slender arms and body, the forces are given by Cox \cite{cox}
slender body theory: The element of force, $d\mbf{F}(s)$, acting
on a segment of length $ds$ located at the point $s$ on the
slender body is given by
    \begin{equation}\label{eq:cox}
    d\mbf{F}(s)=k\big(\mbf{ t} \big( \mbf{t}\cdot \mbf{v}\big) -2
    \mbf{v}\big)\, ds,\quad k=\frac{2\pi\mu}{\ln\kappa}
    \end{equation}
where $\mbf{ t}(s)$ is a unit tangent vector to the slender-body
at $s$ and $\mbf{v}(s)$ its velocity there. $\mu$ is the viscosity
and $\kappa$ the slenderness (the ratio of length to diameter).

The force on the a-th arm depends linearly on the velocities of
the controls $\dot\theta_j$ and swimming velocity $\dot X$. For
example, the force component in the x-direction on the a-th arm
takes the form
    \be
    {F}_{a}^x= f_{aj}^x\,\dot \theta_j+f_{a}^{xx}\,\dot X
    \ee
where $f_{aj}^x$ are functions of the controls, given by
elementary integrals
    \be\label{eq:f_jk}
    f_a^{xx}= k (\cos^2\theta_a-2) \int_0^{\ell}ds\,,
\quad    f_{aj}^x=2 k\sin\theta_j\, \int_0^{\ell}s\, ds
    \ee
Similar equations hold for the left arm and the body. The
requirement that the total force on the swimmer vanishes gives a
linear relation between the variation of the controls and the
displacement $\dbar x$
    \begin{equation}\label{eq:dx}
    \dbar x=\frac{\dbar X}\ell=-a(\xi,\eta)\,(d\xi_1+d\xi_2),\quad \xi_j=\cos\theta_j,
\end{equation}
where
    \be\label{eq:a-def}
    a(\xi,\eta)= \frac{1}{4+\eta- \xi_1^2-\xi_2^2}, \quad
    \eta=\frac {\ell_0}{2\ell}
    \ee
As one expects, the body is just a ``dead weight'' and a trim
swimmer with $\eta=0$ is best.

\subsection{The curvature}

The notation $\dbar x$ stresses that the differential displacement
does not integrate to a function of the controls, $x(\xi )$. This
is the essence of swimming: $x$ fails to return to its original
value with $\xi$. For this reason, swimming is best captured not
by the differential one-form $\dbar x$ but by the differential
two-form $\mathcal{F}= d\dbar x$
\begin{equation}\label{eq:F-Yosi}
\mathcal{F}=d\dbar x=2a^2(\xi,\eta) (\xi_2-\xi_1)\, d\xi_1\wedge
d\xi_2
\end{equation}
and $a(\xi)$ is the rational function given in
Eq.~(\ref{eq:a-def}). $\mathcal{F}$ is commonly known as the
curvature \cite{Wilczek-Shapere} and its surface integral for a
region enclosed by a curve $\gamma$ gives, by Stokes, the distance
covered in one stroke. It gives complete information on both the
direction of swimming and distance.

The total curvature associated with the full square of shape
space, $|\xi_j|\le 1$ is 0 by symmetry. The total positive
curvature associated with the triangular half of the square,
$\xi_1\le \xi_2$ is 0.274. This means that swimmer can swim, at
most, about a quarter of its arms length in a single stroke.

$\mathcal{F}$ is a differential two form and as such is assigned a
numerical value only when in comes with a region of integration.
By itself, it has no numerical value.  To say that the curvature
is large at a given point in shape, or control, space requires
fixing some a-priori measure. For example one can pick the flat
measure for $\xi$ in which case $a^2(\xi,\eta )(\xi_1-\xi_2)$
gives numerical values for the curvature. However, one can pick
instead the flat measure for $\theta$ one gets a different
function. A natural measure on shape space is determined by the
dissipation. We turn to it now.

\subsection{The metric in shape space}

The power of swimming at low Reynolds numbers is
quadratic in the driving: $g_{jk}\,\dot\theta_j\dot\theta_k$ where
$g_{jk}(\theta)$ is a function on shape space and we use the
summation convention where repeated indices are summed over. This
suggests the natural metric in shape space is, in either
coordinate systems,
    \begin{equation}\label{eq:energy dissipation}
    g_{jk}(\theta)\,d\theta_j d\theta_k=g_{jk}(\xi)\,d\xi_j d\xi_k
    \end{equation}
In particular, the associated area form is
    \be
    \sqrt{\det g(\theta)}\, d\theta_1\wedge d\theta_2=\sqrt{\det
    g(\xi)}\, d\xi_1\wedge d\xi_2
    \ee
The curvature can now be assigned a natural numerical value
    \be\label{eq:numerical-curvature}
    2 a^2(\xi,\eta)\frac{\xi_1-\xi_2}{\sqrt{\det g(\xi)}}=
    2 a^2(\xi,\eta)\frac{\xi_1-\xi_2}{\sqrt{\det
    g(\theta)}}\sin\theta_1\sin\theta_2
    \ee

Each arm of the symmetric Purcell swimmer dissipate energy at the
rate
   \bea\label{eq:dissipation-arm}
    -\int_0^\ell d\mbf{F}(s)\cdot \mbf{v}ds &=&-k\int_0^\ell\big(( \mbf{t}\cdot \mbf{v})^2 -2
    \mbf{v}\cdot \mbf{v}\big)\,ds \\
    &=&-k\int_0^\ell\big( \dot X^2 -
    2 s^2 \dot\theta_j^2  \cos^2\theta_j-2(s \dot\theta_j\,\sin\theta_j-\dot X)^2 \big)\,ds
    \nonumber \\
   &=&\frac{k \ell^3} 3 \left( 3\dot x^2 +
     2  \dot\theta_j^2 +6 \dot\xi_j \dot x\right)\nonumber
    \eea
And the total energy dissipation by the arms is evidently
    \be\label{eq:dissipation}
    \frac{2k \ell^3} 3 \left( 6\dot x^2+
     2  (\dot\theta_1^2+\dot\theta_2^2) +6 (\dot\xi_1+\dot\xi_2) \dot x\right)
    \ee
In a body-less swimmer, $\eta=0$, this is also the total
dissipation, and we consider this case from now on. Since we are
interested in the metric up to units, we shall henceforth
set\footnote{An alternate natural normalization is one that
preserves the area $4\pi^2$.} $4k \ell^3=3$. Plugging the swimming
equation, Eq.~(\ref{eq:dx}) gives $g$:
    \bea\label{eq:Metric}
    g(\theta)=a(\xi,0)
    \left(%
    \begin{array}{cc}
    5 - 2\xi_2^2  + \xi_1^2 &\sin \theta _1 \sin \theta _2 \\
    \sin \theta _1 \sin \theta _2& 5 - 2\xi_1^2  + \xi_2^2  \\
    \end{array}%
    \right),\quad \xi_j=\cos\theta_j
    \eea
and $a(\xi,0)$ is given in Eq.~(\ref{eq:a-def}). In particular, $
g(\theta)$ is a smooth function on shape space while $g(\xi)$ is
singular at the boundaries. One can now meaningfully plot the
curvature which is shown in Fig. \ref{fig:OptStrkHH}

\subsection{The optimal stroke}

Efficient swimming covers the largest distance for given energy
resource and at a given speed\footnote{One needs to constrain the
average speed since one can always make the dissipation
arbitrarily small by swimming more slowly.}. Alternatively, it
minimizes the energy needed  for covering a given distance at a
given speed\footnote{The distance is assumed to be large compared
with a single stroke distance. The number of strokes is not given
a-priori.}. Fixing the speed for a given distance is equivalent to
fixing the time $\tau$. In this formulation the variational
problem takes the form of a problem in Lagrangian mechanics of
minimizing the action
\begin{equation}\label{eq:Action}
\int\limits_{0}\limits^{\tau}  g_{jk}(\theta)\,
\dot\theta_j\dot\theta_k dt +q\int \dbar x
\end{equation}
where q is a Lagrange multiplier.  $\dbar x$ is given in
Eq.~(\ref{eq:dx}). This can be interpreted as a motion of a a
charged particle on a curved surface in an external magnetic field
\cite{Avron-Gat-Kenneth}. Conservation of energy then says that
the solution has constant speed (in the metric $g$). For a closed
path, the kinetic term is then proportional to the length of the
path and the constraint is the flux enclosed by it. Thus the
variational problem can be rephrase geometrically as the
``isoperimetric problem" : Find the shortest path that encloses
the most flux.

\begin{figure}[htb]
  \includegraphics[width=12 cm]{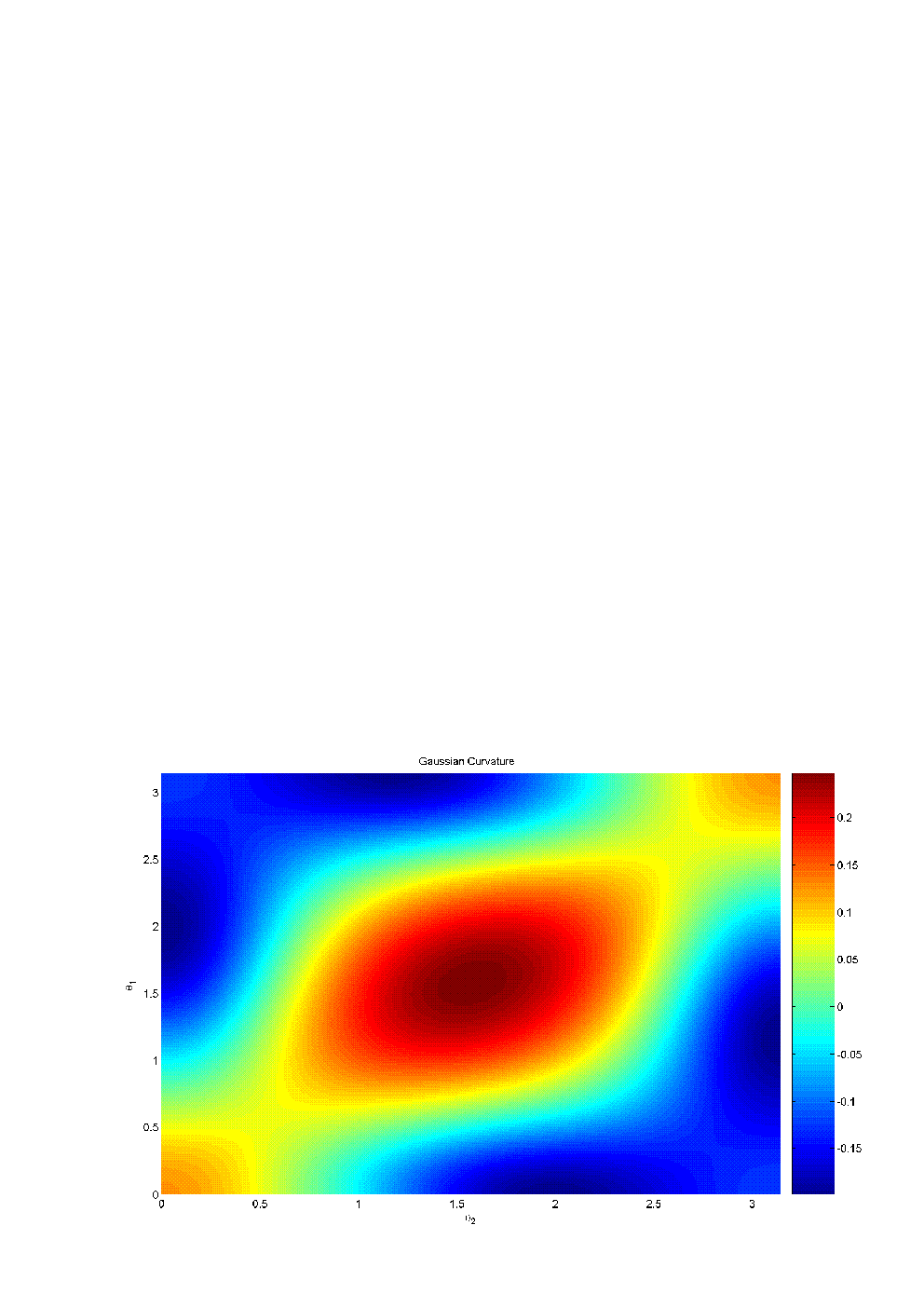}\\
  \caption{The Gaussian curvature on the surface of controls induced by the dissipation metric.
}\label{fig:GaussCurv}
\end{figure}

The charged particle moves on a curved surface. How does this
surface look like? From the dissipation metric we can calculate,
using Brioschi formula \cite{Spivak}, the gaussian curvature $K$
(not to be confused with $\mathcal{F}$) of the surface. A plot of
it is given in Fig.~\ref{fig:GaussCurv}. 

Inspection of Fig.~\ref{fig:OptStrkHH} suggest that pretty good
strokes are those that enclose only one sign of the curvature
$\mathcal{F}$. The actual optimal stroke can only be found
numerically.  It is plotted in figure \ref{fig:OptStrkHH}. The
efficiency for this stroke is about the same as the efficiency of
the Purcell's swimmer for rectangular strokes of \cite{JFM-Stone},
but less than the optimally efficient strokes found in
\cite{Hosoi}.

Cox theory does not allow the  arms to get too close. How close
they are allowed to get depends on the slenderness $\kappa$. The
smallest angle allowed $\delta\theta$ must be such that
$(2\delta\theta) \log \kappa \gg 1$. As the optimal stroke gets
quite close to the boundary, with $\delta \theta\sim 0.1 \ radian$ it
can be taken seriously only for sufficiently slender bodies with
$\log\kappa \gg 5$, which is huge. The optimal stroke is
therefore more of mathematical than physical interest. One can use a refine slender body approximation by taking high order terms in Cox's expansion for the force. This will leave the structure without changes, but will made Eq.~\ref{eq:F-Yosi} and Eq.~\ref{eq:Metric} much more complicated.

From a mathematical point of view it is actually quite remarkable
that a minimizer exists. By this we mean that the optimal stroke
does not hit the boundary of shape space $|\xi_j|=1$ where Cox
theory is squeezed out of existence. This can be seen from the
following argument. Inspection of
Eq.~(\ref{eq:numerical-curvature}) and Eq.~(\ref{eq:Metric}) shows
that the curvature vanishes linearly near the boundary of shape
space (this is most easily seen in the $\theta$ coordinates).
Suppose now that the optimal path ran along the boundary. Shifting
the path a distance $\e$ away from the boundary would shorten it
linearly in $\e$ while the change in the flux integral will be
only quadratic. This shows that the path that hits the boundary
can not be a minimizer.

\section{Purcell swimmer}

Purcell swimmer can move in either direction in the plane and can
also rotate. Since the Euclidean group is not Abelian (rotations
and translations do not commute) the notion of ``swimming
curvature'' that proved to be so useful in the Abelian case needs
to be modified. As we shall explain, landscape figures can be used
to give qualitative geometric understanding of the swimming and in
particular can be used to answer  Purcell question ``What will
determine the direction this swimmer will swim?''. However, unlike
the Abelian case, the swimming curvature does not give full
quantitative information on the swimming and one can not avoid
solving a system of differential equations in this case if one is
interested in quantitative details.

The location and orientation of the swimmer (in the Lab frame)
shall be denoted by the triplet $x^\alpha$ where $x^0=\phi$ is the
orientation of the swimmer, see Fig.~\ref{fig:purcell}, and
$x^{1,2}$ are cartesian coordinates of the center of the ``body''
\footnote{All distances are dimensionless being measured in units
of arm length $\ell$.}. We use super-indices and Greek letters to
designate the response while lower indices and Roman characters
designate the controls $|\theta_j|<\pi$.

\subsection{Linear response versus a gauge theory}

The common approach to low Reynolds numbers swimming is to write
the equations of motion in a fixed, lab frame. We first review
this and then describe an alternate  approach where the equations
of motions are written in a frame that instantaneously coincides
with the swimmer.

 By general principles of low Reynolds numbers
hydrodynamics there is a linear relations between the change in
the controls\footnote{In this section we use the convention the
$\theta_1$ increases counterclockwise and $\theta_2$ clockwise.}
$d\theta$ and the response $dx$
    \be\label{eq:linear-response}
    \dbar x^\alpha={\mathcal A}^\alpha_j \, d\theta_j\,,
    \ee
(summation over repeated indices implied.) Note that $j=1,2$ since
there are two controls, while $\alpha=0,1,2$ for the three
responses.

The response coefficients $\mathcal{A}$ are functions of both the
control coordinates $\theta_k$ and the location coordinates
$x^\beta$ of the swimmer in the Lab. However, in a homogeneous
medium it is clear that ${\mathcal A}^\alpha_j$ can only be a
function of the orientation $x^0=\phi$. Moreover, in an isotropic
medium it can only dependent on the orientation $\phi$ through
    \be\label{eq:RA}
    \mathcal{A}^\alpha_i \left( {\phi ,\theta  } \right) =
    R^{\alpha\beta}(\phi) A^\beta_i \left( {\theta  }
    \right); \quad R^{\alpha\beta}\left( \phi \right) = \left(
    {\begin{array}{*{20}c}
    1 & 0 & 0 \\
   0 &\cos \phi  &  - \sin \phi   \\
   0 & \sin \phi  & \cos \phi    \\
    \end{array}} \right)
    \ee
In the Lab frame, the nature of the solution of the differential
equations is obscured by the fact that one can not determine $dx$
independently for different points on the stroke (because of the
dependence on $\phi=x^0$).

    The coefficients $A^\beta_i$ may be viewed as the transport
coefficients in a rest frame that instantaneously coincides with
the swimmer. They play a key role in the geometric picture that we
shall now describe. In the frame of the swimmer one has
 \be\label{eq:linear-response-non-ab}
    \dbar y^\alpha= A^\alpha_j \, d\theta_j
    \ee
which is an equation that is  fully  determined by the controls.
The price one pays is that the $\dbar y$ coordinates cannot be simply added to calculate the total change in a stroke $\gamma$, since one has to consider the changes in the reference frame as well. In order to do that, $\dbar y$  must be viewed as (infinitesimal) elements of the Euclidean group
    \be\label{eq:euclidean}
    E(y^\alpha)= \left(%
    \begin{array}{ccc}
     \cos y^0 & \sin y^0 & y^1 \\
    - \sin y^0& \cos y^0 & y^2 \\
     0 & 0 & 1 \\
    \end{array}%
    \right)
    \ee
The composition of $\dbar y$ along a stroke $\gamma$ is a matrix
multiplication
    \be\label{eq:matrix-prod}
    E(\gamma)= \prod_{\theta\in \gamma} E\big(dy^\alpha(\theta)\big)
    \ee
The product is, of course, non commutative. We denote generators
of translations and rotations by
 \be\label{eq:translations}
 e^0= \left(%
    \begin{array}{ccc}
     0 & 1  & 0 \\
     -1& 0 & 0 \\
     0 & 0 & 0 \\
    \end{array}%
    \right),\quad
    e^1= \left(%
    \begin{array}{ccc}
     0 & 0  & 1 \\
     0& 0 & 0 \\
     0 & 0 & 0 \\
    \end{array}%
    \right),\quad
    e^2= \left(%
    \begin{array}{ccc}
     0 & 0  & 0 \\
     0& 0 & 1 \\
     0 & 0 & 0 \\
    \end{array}%
    \right)
    \ee
They satisfy the Lie algebra
    \be
    [e^0,e^\alpha]=-\e^{0\alpha \beta }e^\beta,\quad  [e^1,e^2]=0
    \ee
where $\varepsilon^{\alpha \beta \gamma}$ is the completely
anti-symmetric tensor. One can write
Eq.~(\ref{eq:linear-response-non-ab}) concisely as a matrix
equation
    \be\label{eq:matrix-form}
    \dbar y =A_j d\theta_j,\quad \dbar y= y^\alpha e^\alpha, \quad
    A_j=A_j^\alpha e^\alpha
    \ee
where $\dbar y$ and $A_j$ are $3\times 3$ matrices (summation over
repeated indices implied).

\begin{figure}[htb]
  \includegraphics[width=12cm]{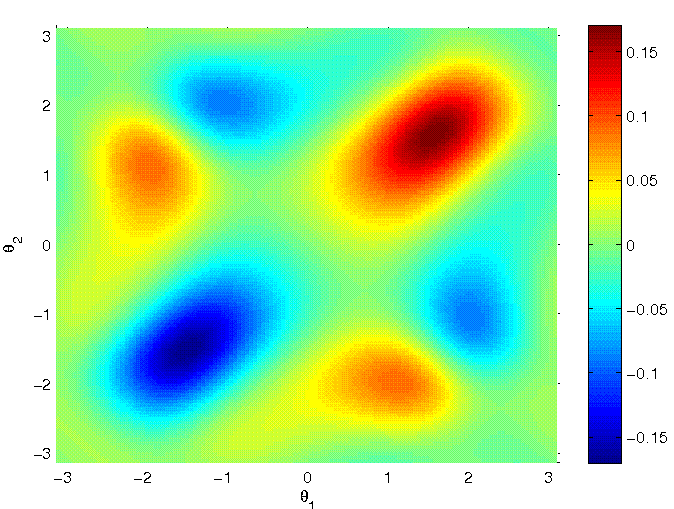}\\
  \caption{The curvature for rotation 
  for Purcell's three linked swimmer with $\eta=2$, plotted with the flat measure on $(\theta_1,\theta_2)$}\label{fig:FphiEta=2}
\end{figure}

\begin{figure}[htb]
  \includegraphics[width=10cm]{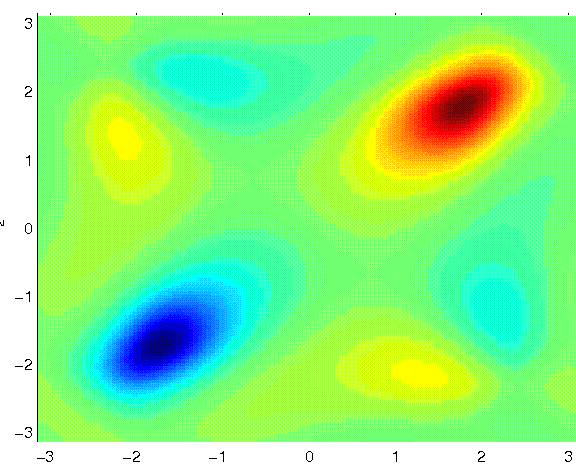}\\
  \caption{The rotation curvature 
  for Purcell's three linked swimmer with $\eta=2$ plotted using the measure induced by dissipation}\label{fig:FphiNorm}
\end{figure}

\subsection{The swimming curvatures}\label{sec:curvature}

Once the (six) transport coefficients $A^\alpha_j$ are known, one
can, in principle, simply integrate the system of three, first
order, non-linear ordinary differential equations,
Eq.~(\ref{eq:linear-response}). This can normally be done only
numerically. Numerical integration is practical and useful, but
not directly insightful. We want to describe tools that allow for
a qualitative understating swimming in the plane without actually
solving any differential equation.

Low Reynolds numbers swimmers perform lots of mutually cancelling
maneuvers with a small net effect. The swimming curvature measure
only what fails to cancel for infinitesimal strokes. Since
reversing a loop reverses the response, it is natural to expect,
that $\delta y^\alpha$ for a closed (square) loop is proportional
to the area form. Integrating  Eq.~(\ref{eq:matrix-form}) around a
closed infinitesimal loop gives
    \be
     \delta y   =\mathcal{F} d\theta_1\wedge d\theta_2
    \ee
where
     \be
     \mathcal{F}=
    \partial_1 A_2-
    \partial_2 A_1
    -[A_1,A_2], \quad \mathcal{F}= \mathcal{F}^\alpha e^\alpha,\quad \quad \partial_j=\frac{\partial}{\partial \theta_j}
    \ee
$\mathcal{F}$ and $\delta y$ are $3\times 3$ matrices.
$\mathcal{F}$ has the structure of curvature of a non-abelian
gauge field \cite{nakahara}.  In coordinates, this reads
    \be
     \mathcal{F}^\alpha=
    \partial_1 A^\alpha_2-
    \partial_2 A^\alpha_1
    +\varepsilon^{0\alpha\beta }\left( A^0_1  A^\beta_2
    - A^0_2  A^\beta_1 \right),
    \ee

In the Lab coordinates one has, of course,
     \be
    \delta x^\alpha=\mathcal{\tilde{F}}^\alpha d\theta_1\wedge d\theta_2,\quad
       \mathcal{\tilde{F}}^\alpha= R^{\alpha\beta}(\phi) \mathcal{F}^\beta(\theta),
    \ee
The curvature is Abelian when the commutator vanishes. This is the
case in  Eq.~(\ref{eq:numerical-curvature}) and it is also the
case for the rotational curvature. The Abelian curvature gives
full information on the swimming of finite stroke by simple
application of Stokes formula.  This is, unfortunately, not the
case in the non-Abelian case. One can not reconstruct the
translational motion of a large stroke from the infinitesimal
closed strokes $\delta y$ because Stokes theorem only works for
commutative coordinates and $\delta y$ are not.

\begin{figure}[htb]
  \includegraphics[width=8cm]{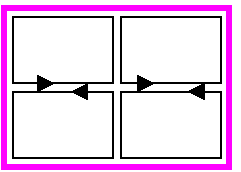}\\
  \caption{The failure of Stokes theorem in the non-commutative case: The integration
  on the adjacent segments traversed in opposite senses do not cancel in the non-commutative case.
  }\label{fig:stokes}
\end{figure}

For Purcell swimmer, $\mathcal{F}$ although explicit, is rather
complicated. Since the dissipation metric is complicated too, we
give two plots of $\mathcal{F}$:
Figs.~(\ref{fig:FphiEta=2}, \ref{fig:PurcellCurvX},\ref{fig:PurcellCurvY}) give the
curvature relative to the flat measure on $(\theta_1,\theta_2)$,
and describe how far the swimmer swims for small strokes. In
Figs.~(\ref{fig:FphiNorm},~\ref{fig:PurcellCurvXNorm},\ref{fig:PurcellFyNorm}) the
curvature is plotted relative to the dissipation measure and it
displays the energy efficiency of strokes.

\subsection{The equations of motion}
Using Cox theory, in a manner analogous to what was done in for
the symmetric swimmer, one can calculate explicitly the force (and
torque) on the $a$-th rod in the form
    \be
    F_a^\alpha= f_{aj}^\alpha\dot\theta_j +f_a^{\beta\alpha}\dot
    x^\beta
    \ee
where $f_{aj}^\alpha$ are explicit and relatively simple functions
of the controls (compare with Eq.~(\ref{eq:f_jk})). The swimming
equation are then given by
    \be
   \sum_a F_a^\alpha=\left( \sum_af_{aj}^\alpha\right)\dot\theta_j +
   \left(\sum_a f_{a}^{\alpha\beta}\right)
   \dot
    x^\beta=0
    \ee
This reduces the problem of finding the connections $A$ to a
problem in linear algebra. Formally
    \be
    {A}^\beta_j=\left(\sum_a f_a^{\beta\alpha}\right)^{-1}\left( \sum_af_{aj}^\alpha\right)
    \ee
where the bracket on the left is interpreted as a $3\times 3$
matrix, with entries $\alpha,\beta$, and the inverse means an
inverse in the sense of matrices. Although this is an inverse of
only a $3 \times 3$ matrix the resulting expressions are not very
insightful. We spare the reader this ugliness which is best done
using a computer program.

\subsection{Symmetries}
Picking the center point of the body as the reference fiducial
point is, in the terminology of Wilczek and Shapere
\cite{Wilczek-Shapere} a choice of gauge. This particular choice
is nice because it implies symmetries of the connection $A$
\cite{Hosoi,JFM-Stone}.  Observe first that the
interchange $(\theta_1,\theta_2)\rightarrow (-\theta_2,-\theta_1)$
corresponds to a rotation of the swimmer by $\pi$. Plugging this
in Eq.~(\ref{eq:RA}) one finds
    \be\label{eq:symmetries-r}
    A^\beta_1 ( \theta_1,\theta_2)=\left\{%
\begin{array}{ll}
    +A^\beta_2 ( -\theta_2,-\theta_1), & \hbox{$\beta =0$;} \\
    -A^\beta_2 ( -\theta_2,-\theta_1), & \hbox{otherwise.} \\
\end{array}%
\right.
    \ee
This relates the two half of the square divided by the diagonal
$\theta_1+\theta_2=0$.

A second symmetry comes from the interchange
$(\theta_1,\theta_2)\rightarrow (\theta_2,\theta_1)$ corresponding
to the reflection of the swimmer around the central vertical of
the middle link. Some reflection shows then that
    \be\label{eq:symmetries-2}
    A^\beta_1 ( \theta_1,\theta_2)=\left\{%
\begin{array}{ll}
    +A^\beta_2 ( \theta_2,\theta_1), & \hbox{$\beta =2$;} \\
    -A^\beta_2 ( \theta_2,\theta_1), & \hbox{otherwise.} \\
\end{array}%
\right.
    \ee
This relates the two halves of  the square divided by the diagonal
$\theta_1=\theta_2$.

 The symmetries can be combined to yield the result
that $A^0$ and $A^2$ are anti-symmetric  and $A^1$ is symmetric
under inversion
    \be
    A^0_j(\theta)=-A^0_j(-\theta), \quad A^1_j(\theta)=A^1_j(-\theta), \quad A^2_j(\theta)=-A^2_j(-\theta)
    \ee

\subsubsection{Rotations}

The rotational motion of Purcell swimmer, in any finite stroke, is
fully captured by the Abelian curvature
     \be
    \mathcal \mathcal{F}^0 = \mathcal{F}^0=\partial_1 {A}^0_2-
    \partial_2 {A}^0_1
    \ee
This reflects the fact that rotations in the plane are
commutative.

The symmetry of Eq.~(\ref{eq:symmetries-r}) implies
  \be\label{eq:symmetries-r-d}
   ( \partial_2A^0_1 )( \theta_1,\theta_2)=
    (\partial_1A^0_2) ( -\theta_2,-\theta_1)
    \ee
and this says that  $ \mathcal{F}^0$is \emph{ani-symmetric} under
reflection in the diagonal $\theta_1+\theta_2=0$. Similarly,
Eq.~(\ref{eq:symmetries-2}) implies \be\label{eq:symmetries-2}
    (\partial_2A^0_1) ( \theta_1,\theta_2)=
    -(\partial_1A^0_2) ( \theta_2,\theta_1),
    \ee
and this says that $\mathcal{F}^0$ is  \emph{symmetric} about the
line $\theta_1=\theta_2$. Fig.~(\ref{fig:FphiEta=2}) is a plot of
the curvature and it clearly has the requisite symmetries.

The total curvature associated with the full square of shape space
vanishes (by symmetry). For $\eta=2$, one can see in Fig.~\ref{fig:FphiEta=2} three positive islands
surrounded by three negative lakes. The total curvature associated
with the three islands is quite small, about 0.1. This
means that Purcell swimmer with $\eta=2$ turns only a small
fraction of a circle in any full stroke.

\subsection{Translation}

\begin{figure}[htb]
  \includegraphics[width=14cm]{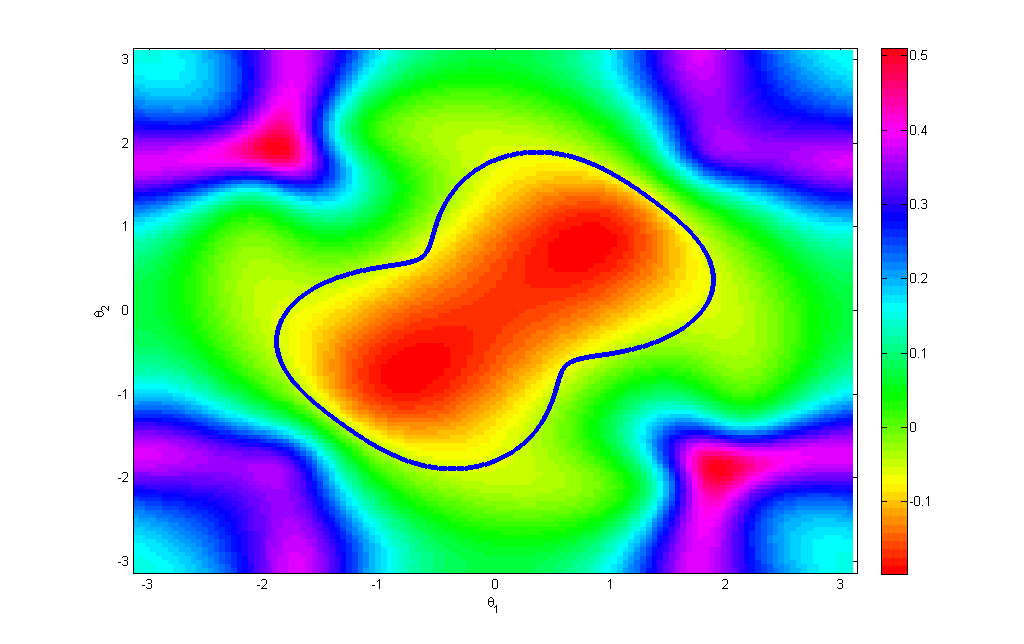}\\
  \caption{The landscape of the x-curvature for Purcell's swimmer with $\eta=0.75$
  shown with the the Tam and Hosoi optimal distance stroke \cite{Hosoi}. The curvature is given  relative to the flat
  measure in  $(\theta_1,\theta_2)$.  }\label{fig:PurcellCurvX}
\end{figure}


\begin{figure}[htb]
\includegraphics[width=12cm]{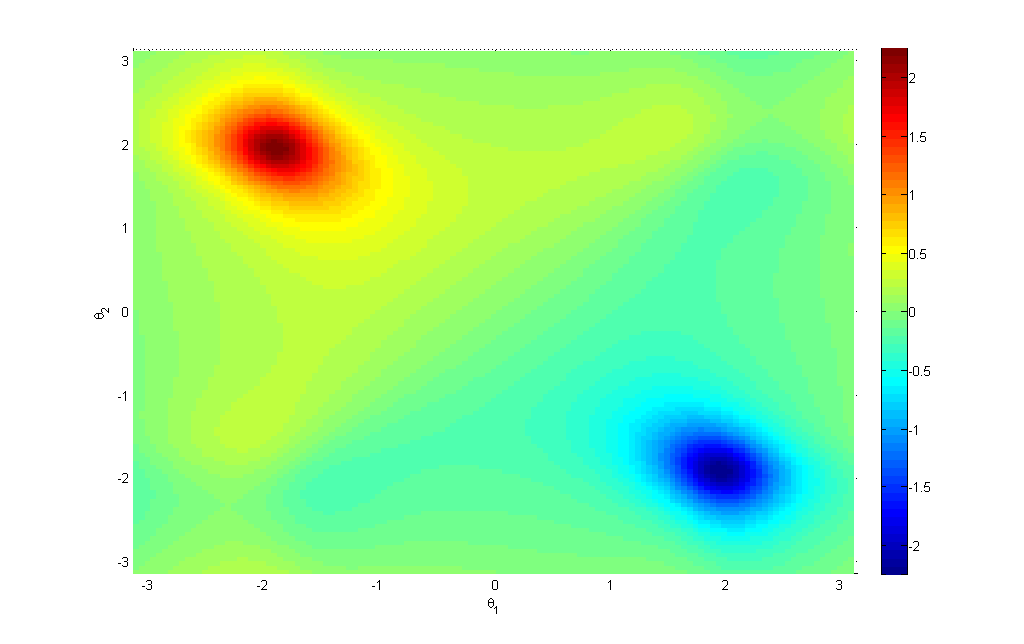}
\caption{The landscape of the y-curvature for Purcell's swimmer with $\eta=0.75$. The curvature is given  relative to the flat measure in  $(\theta_1,\theta_2)$.  }\label{fig:PurcellCurvY}
\end{figure}

The curvatures corresponding to the two translations of a swimmer with $\eta=0.75$
are shown in Figs.~(\ref{fig:PurcellCurvX},\ref{fig:PurcellCurvY},\ref{fig:PurcellCurvXNorm},\ref{fig:PurcellFyNorm})(here we use $\eta=0.75$ for comparison with \cite{Hosoi}). The symmetries of the figures are a consequence of
Eqs.~(\ref{eq:symmetries-r},\ref{eq:symmetries-2}). Form the first
we have \be\label{eq:symmetries-3}
   ( \partial_2A^\beta_1 )( \theta_1,\theta_2)=-
    (\partial_1A^\beta_2) ( -\theta_2,-\theta_1), \beta=1,2
    \ee
which implies that  $ \mathcal{F}^{1,2}$ are \emph{symmetric}
under reflection in the diagonal $\theta_1+\theta_2=0$. Similarly,
from the Eq.~(\ref{eq:symmetries-2}) we have
    \be\label{eq:symmetries-4}
    (\partial _2A^\beta_1) ( \theta_1,\theta_2)=\left\{%
\begin{array}{ll}
    +(\partial _1A^2_2) ( \theta_2,\theta_1), & \hbox{$\beta =2$;} \\
    -(\partial_1A^1_2) ( \theta_2,\theta_1), & \hbox{$\beta =1$.} \\
\end{array}%
\right.
    \ee
This says that $ \mathcal{F}^{1}$ \emph{symmetric} and  $
\mathcal{F}^{2}$ \emph{anti-symmetric}  under reflection in the
diagonal $\theta_1=\theta_2$.

The curvatures for the translations is non-Abelian and can not be
used to {\em calculate} the swimming distances for {\em finite}
strokes because the Stokes theorem fails.

\section{Qualitative analysis of swimming}

\subsection{When is a stroke small?}
The landscape figures for the translational curvatures provide
precise information on the swimming distance for infinitesimal
strokes. They are then also useful to characterize small strokes.
The question is how small is small? For a stroke of size $\e$, the
controls are of size $\delta\theta=O(\e)$ and the swimming
distance measures by the curvature is $O(\mathcal {F}\e ^2)$. The
error in this has terms\footnote{There are also terms of the form
the form $O( A^3 \e^3)$.} of the form $O(A\mathcal{F} \e^3)$. This
suggest that the relative error in the swimming distance as
measured a finite stroke is of the order $O(A\e )$. Hence, a
stroke is small provided $|A\e| \ll 1$. Clearly, a Purcell swimmer
swims substantially less than an arm length as the arm moves. This
says that $|A|\ll 1$ and so strokes of the order of a radian can
be viewed as small strokes.

A radian is the scale of the structures in the landscape of the
figures of the curvature. This means that the landscape carries
qualitative information about the swimming of moderate strokes.

\subsection{x versus y}

The x-curvature is symmetric under inversion
    \be
     \mathcal{F}^1(\theta)= \mathcal{F}^1(-\theta)
    \ee
Since both $A^0$ and $A^2$ are antisymmetric under inversion, one
sees that the non-Abelian part of the x-curvature is of order
$O(\theta^2)$ near the origin. The x-translational curvature,
which is non-zero near the origin, is almost Abelian for small
strokes.

The y-curvature, in contrast, is anti-symmetric under inversion
    \be
     \mathcal{F}^12(\theta)= -\mathcal{F}^2(-\theta)
    \ee
and so vanishes linearly at the origin. The non-Abelian part is
also anti-symmetric under inversion, and it too vanishes linearly.
The y-curvature is therefore {\em not} approximately Abelian for
small strokes, but it is small.

\subsubsection{Which way does a swimmer swim?} The swimming direction can be easily determined
for those strokes that live in a region where the translational
curvature has a fixed sign. This answers Purcell's question for
many strokes. Strokes the enclose both signs of the curvature are
subtle.

\subsubsection{Subtle swimmers}
Purcell's swimmer can reverse its direction of propagation by
increasing the stroke amplitude \cite{JFM-Stone}. This can be seen
from the landscape diagram, Fig.~(\ref{fig:PurcellCurvX}): Small
square strokes near the origin  sample only slightly negative
curvature. As the stroke amplitude increases the square gets
larger and begins to sample regions where the curvature has the
opposite sign, eventually sampling regions with substantial
positive curvature.

\subsubsection{Optimal distance strokes}
The curvature landscapes are useful when one wants to search for optimal strokes as they
provide an initial guess for the stroke. (This initial guess can
then be improved by standard optimization numerical methods.)

For example, Tam and Hosi \cite{Hosoi} looked for strokes that
cover the largest possible distance. For strokes near the origin,
a local optimizer is the stroke that bounds the approximate square
blue region in Fig.~\ref{fig:PurcellCurvX} (in this case $\eta=0.75$).

\begin{figure}[htb]
  \includegraphics[width=12cm]{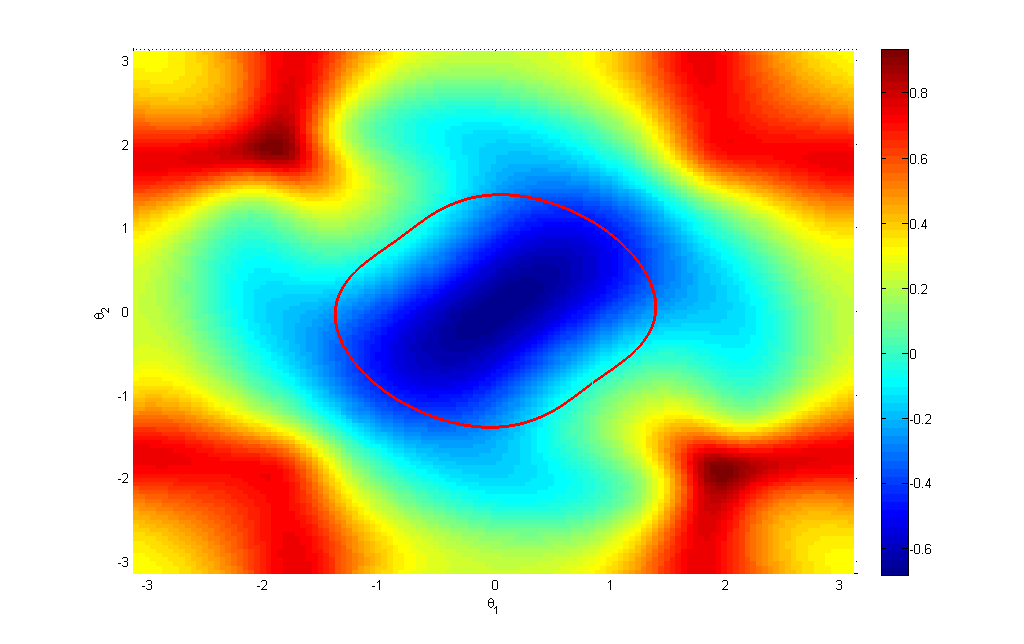}\\
  \caption{The x-curvature relative to the dissipation metric
  and the optimal efficient stroke found by Tam and Hosoi \cite{Hosoi}.}\label{fig:PurcellCurvXNorm}
\end{figure}

\begin{figure}[htb]
  \includegraphics[width=12cm]{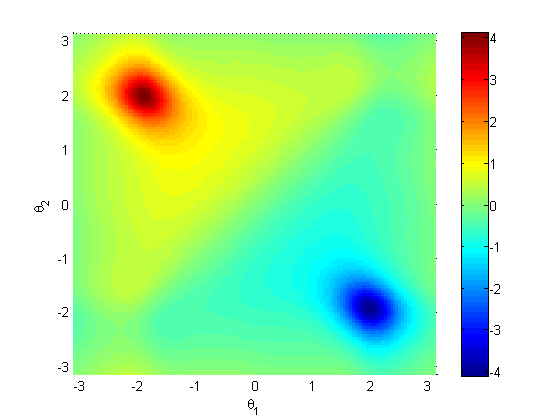}\\
  \caption{The y-curvature relative to the dissipation metric. Pay attention to the large values in the scale.}\label{fig:PurcellFyNorm}
\end{figure}

\begin{figure}
  \includegraphics[width=12 cm]{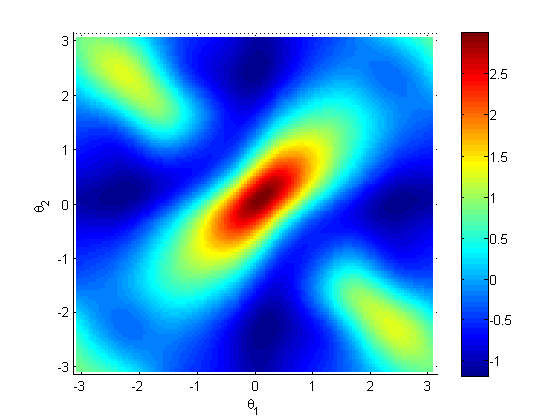}\\
  \caption{The gaussian curvature induced by the dissipation metric of Purcell's swimmer}\label{fig:GaussCurvPurcell}
\end{figure}

\subsubsection{Efficient strokes}

The curvature normalized by dissipation, Fig.
\ref{fig:PurcellCurvXNorm} gives a guide for finding efficient small
strokes. Caution must be made, since while the displacement can be approximated from the surface area, the energy dissipation is proportional to the stroke's length and not the stroke's area. In regimes where the Gaussian curvature of the dissipation (Fig.~\ref{fig:GaussCurvPurcell}) is positive - it is possible to have strokes with small length which bounds large area. In the case of Purcell's swimmer, this suggest two possible regimes: around the origin and the positive curvature island in the upper left (lower right) corner of Fig.~\ref{fig:GaussCurvPurcell}. The optimizer near the origin is the optimal stroke found in \cite{Hosoi}, while the optimizer in the upper left corner \cite{Comment} - although more efficient (pay attention to the values of $\frac{\mathcal F^y}{\sqrt{\det g(\theta)}}$ in Fig.~\ref{fig:PurcellFyNorm}), is of mathematical interest only, since it is near the boundary, where the first order slender body approximation Eq.(~\ref{eq:cox}) is relevant only for extremely slender bodies.



\paragraph{ Acknowledgment} We thank A. Leshanskey and O. Kenneth for
discussions.



\end{document}